\documentclass[prb,twocolumn,showpacs, amsmath, amssymb]{revtex4}
\allowdisplaybreaks
\usepackage{epsfig}
\usepackage{graphicx}

\newcommand{\beq} {\begin{equation}}
\newcommand{\eeq} {\end{equation}}
\newcommand{\beqa} {\begin{eqnarray}}
\newcommand{\eeqa} {\end{eqnarray}}

\begin{document}
\title{The Importance of DNA Repair in Tumor Suppression}
\author{Yisroel Brumer and Eugene I. Shakhnovich}
\affiliation{Harvard University, 12 Oxford Street, Cambridge, Massachusetts 02138}
\date{\today}
\begin{abstract}
The transition from a normal to cancerous cell requires a number of highly 
specific mutations that affect cell cycle regulation, apoptosis,  
differentiation, and many other cell functions. One hallmark of cancerous 
genomes is genomic instability, with
mutation rates far greater than those of normal cells. In microsatellite 
instability (MIN tumors), these
are often caused by damage to 
mismatch repair genes, allowing further mutation of the genome and tumor 
progression. These mutation rates may lie near the 
error catastrophe found in the quasispecies model of adaptive RNA genomes, 
suggesting that further increasing mutation rates will destroy cancerous 
genomes. However, recent results have demonstrated that  DNA genomes exhibit 
an error threshold at mutation rates far lower than their conservative 
counterparts.
Furthermore, while the maximum viable mutation rate in conservative systems 
increases indefinitely with increasing master sequence fitness, the semiconservative threshold
plateaus at a relatively low value. This implies a paradox,
wherein inaccessible mutation rates are found in viable tumor
cells. In this paper, we address this paradox, demonstrating an isomorphism 
between the conservatively replicating (RNA) quasispecies model and the
semiconservative (DNA) model with post-methylation DNA repair mechanisms
impaired. Thus, as DNA repair becomes inactivated, the maximum viable 
mutation rate increases smoothly to that of a conservatively replicating 
system on a transformed landscape, with an upper bound that is dependent on 
replication rates. On a specific single fitness peak landscape, the
repair-free semiconservative system is shown to mimic a conservative 
system exactly .
We 
postulate that inactivation of post-methylation repair mechanisms are 
fundamental to the progression of a tumor cell and 
hence these mechanisms act as a method for prevention and destruction of 
cancerous genomes.

\end{abstract}
\pacs{87.14Gg, 87.17.-d, 87.23.-n}
\maketitle

\section{Introduction}

Cancer has presented itself as one of the most difficult challenges science
has ever faced. The complexity of the disease, experimental obstacles, and the
vast array of tumor types have made characterization of the many facets
of tumor progression a slow process. It is now understood that this 
progression requires the alteration of numerous genes, as a genome progresses
from its normal state to a full-blown cancer cell \cite{CancerBook}. 
One important aspect of the cancerous genome lies in its genetic instability.
All cancerous genomes display either high mutation rates (in MIN tumors) or
chromosomal instability (in CIN tumors) \cite{Michor}.

One of the most successful theoretical methods for studying genomic evolution
at high mutation rates has been Eigen's quasispecies model \cite{Eigen}. This
model considers an explicit population of genomes, each made up of $L$ nucleotides
chosen from an alphabet of size $S$, usually chosen to be two for simplicity
or four to model the nucleotides in nature. These genomes replicate, mutate 
and compete on a chosen fitness landscape, a unique mapping of genotype to 
fitness. This is often accomplished by assigning different replication rates 
to each
possible genome and setting all death rates to be equal. The model has yielded a number of impressive and
experimentally verified predictions \cite{Crotty, Kamp, Kamp2, Brumer1} and has recently been used as the
basis for novel anti-viral therapies \cite{Crotty, Loeb}. The main prediction lies in the
idea of an error catastrophe. Below a threshold mutation rate, dubbed the ``error threshold'', the population
evolves, independent of starting conditions, to a distribution of genomes
near the sequence of maximal fitness, often called the master sequence. Above
the threshold mutation rate or the ``error threshold'', the population reaches 
a random distribution with no discernible master. This crossover is
depicted in Fig. 1. These ideas provide a 
method for destroying RNA-based viral genomes. Viruses
are expected to evolve a mutation rate slightly below the error threshold so as
to maintain the capacity to rapidly adapt without surpassing the error 
threshold and becoming inviable. Hence, by increasing the mutation rate of
the species, the virus can be destroyed, and this technique has been 
successfully applied \cite{Crotty}. 
\begin{figure}[ht]
\begin{center}
\includegraphics[width=3.3in]{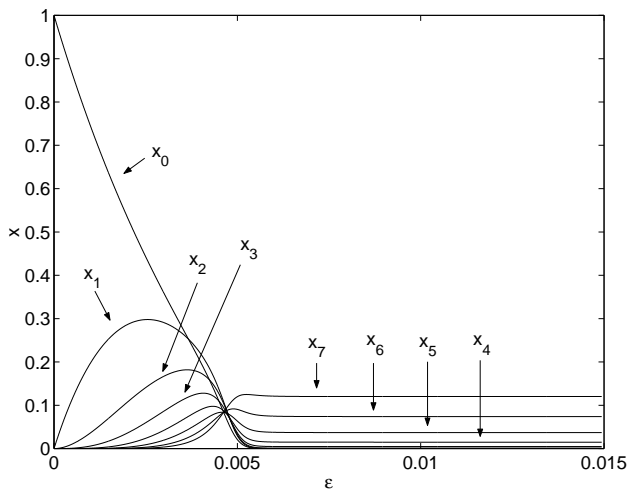}
\end{center}
\caption{A schematic diagram illustrating the error catastrophe predicted by
the quasispecies model. The concentration of sequences of Hamming distance $i$
from the master sequence is represented by $x_i$ and plotted against the
mutation rate, $\epsilon$. At low concentrations, the master sequence (of 
Hamming distance 0), dominates the population, but is surrounded by a cloud 
of closely related genomes . Above the error threshold, this clustering
disappears, and we see a random distribution of genomes, where each Hamming class has a concentration proportional to its size.}
\end{figure}

These ideas have recently been suggested to apply to cancer cells \cite{Sole1}. Cancer and
RNA viruses share genetic instability in the sense that both are rapidly 
mutating and recent work has focused on the idea that mutagens may push
cancer cells past the error threshold in a similar manner. Support for the
idea that the quasispecies model can be applied to complex cellular genomes
comes from recent studies that yielded accurate qualitative and quantitative 
predictions for complex systems such as the adaptive immune system \cite{Kamp}.

However, past work on the quasispecies model has focused on conservatively 
replicating systems such as RNA \cite{Galluccio, Schuster, Tarazona, Peliti, 
timedep, timedep2}. In these systems, single stranded genomes
are copied to produce a new, possibly error prone, strand without affecting the
original. In semiconservative systems like DNA, double stranded genomes
unzip to produce two single strands, each of which is copied to 
produce a new complementary strand by Watson-Crick base-pairing. A variety
of mismatch repair enzymes then repair any errors in the new strand, keeping the effective
error rate low. A few errors remain, though, and these, as well as extrinsic 
mutations induced by UV radiation or other mutagens, are repaired by 
post-methylation repair enzymes that cannot distinguish between the new and old
strands. Thus, some of these base pair mismatches are repaired in the old strand
and the original strand is not conserved \cite{Voet} as shown schematically in
Fig. 2. 

\begin{figure}[ht]
(a)
\begin{center}
\includegraphics[width=2in]{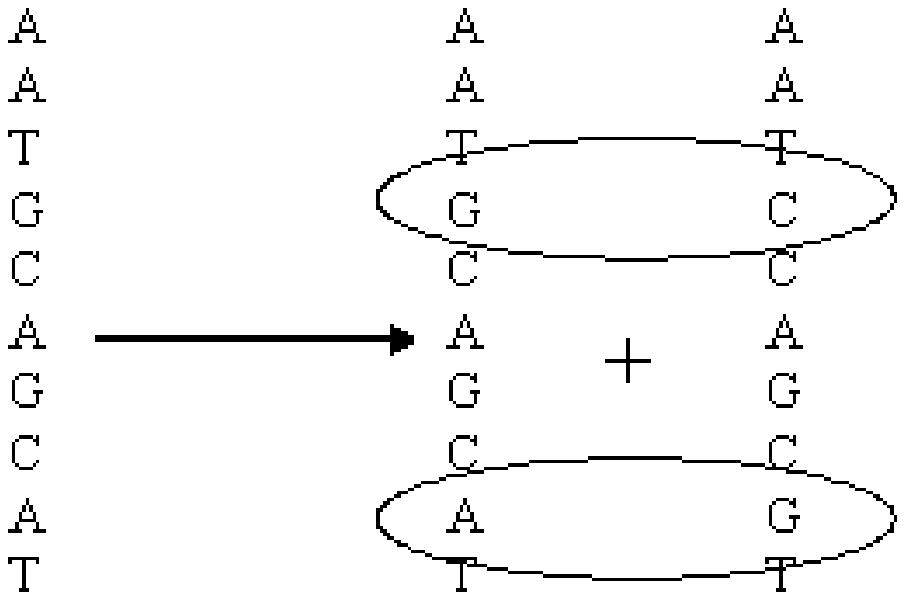}
\end{center}
(b)
\begin{center}
\includegraphics[width=3in]{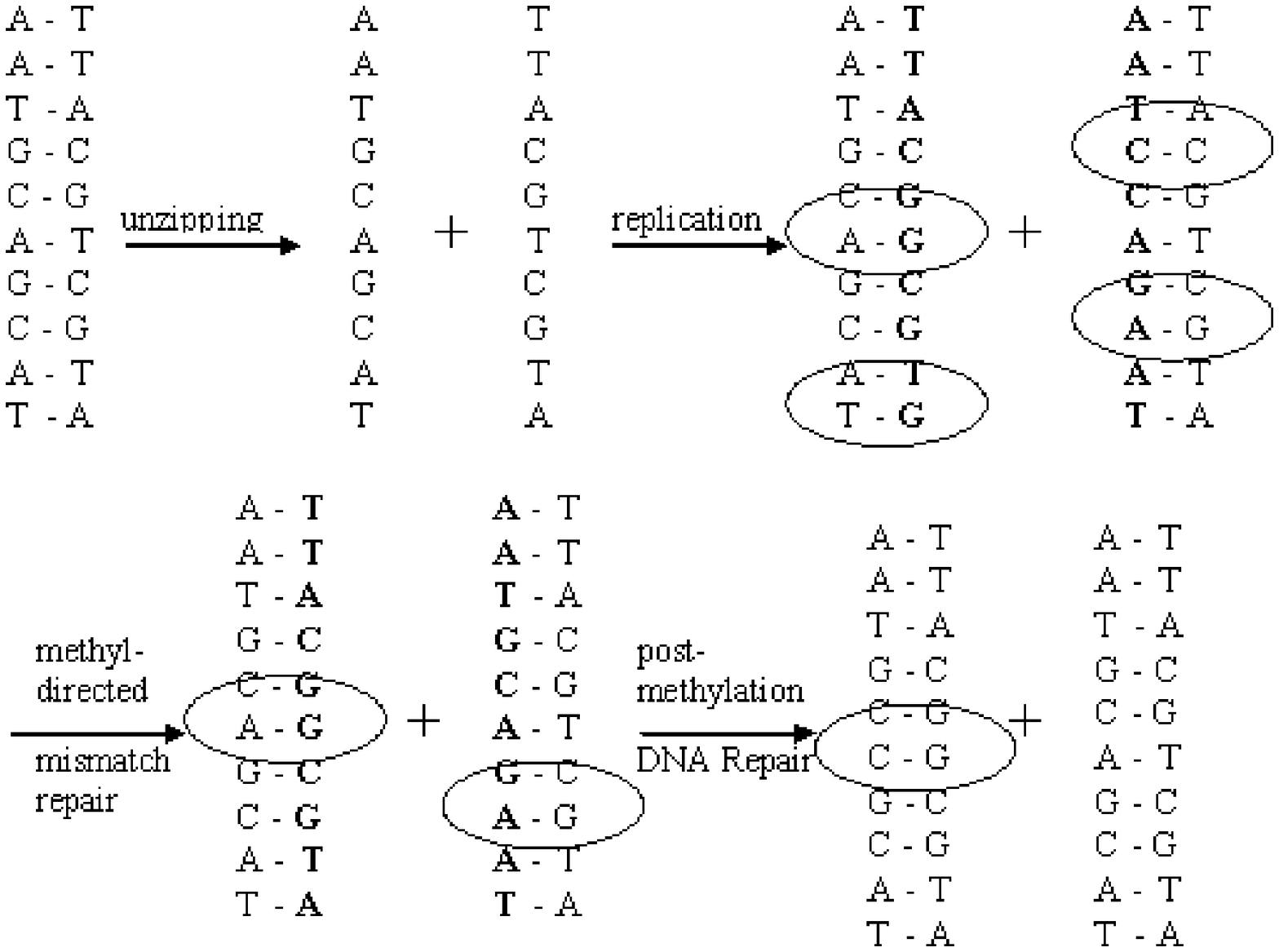}
\end{center}
\caption{A schematic model of (a) conservative and (b) semiconservative 
replication. Non-methylated strands are bolded and errors are circled.}
\end{figure}
The quasispecies model has recently been extended to incorporate this behavior
\cite{Manny}. It was found that, contrary to popular assumption, 
the semiconservative system displays fundamentally different behavior than
the conservative system. In particular, on a single fitness peak landscape,
the semiconservative error catastrophe occurs at significantly lower mutation
rates than the conservative case \cite{Manny}. Particularly interesting is that,
in the conservative case,
the maximum viable mutation rate increases without bounds with increasing 
master sequence replication rate, while 
the semiconservative system reaches a threshold value \cite{Manny}, and this has
been confirmed by simulation for finite populations sizes and genome lengths. Thus,
for the conservative case, it is always possible to ``out-replicate'' the error
threshold. That is, for any given mutation rate, there exists a relative fitness
for the master sequence such that, if the master sequence has that or greater
fitness, the error catastrophe is avoided. This is not true for semiconservative
genomes, where there exist mutation rates that cause the error catastrophe for
{\em any value of the master sequence fitness}. We note in passing 
that a conservative system will also follow this behavior if the possibility
of unrepaired extrinsic mutation is incorporated. The fundamentally different
nature of the semiconservative 
error catastrophe has numerous implications
\cite{Brumer1}, but is particularly pertinent to the study of cancer.

Within the conservative paradigm, it is reasonable to assume that cancer cells 
are capable of maintaining a viable population of rapidly mutating genomes, as
rapid replication rates are one of the hallmarks of cancerous cells 
\cite{CancerBook}. However, the recent results on the semiconservative system 
present a paradox. The mutation rates in MIN cancer cells, known to be 50-1000 times
higher than those of normal cells \cite{Nowak, Cahill}, certainly lie higher than any 
reasonable value for the low semiconservative threshold (for example, the 
single fitness peak landscape yields 1.39 errors/genome/replication as an upper bound for the error threshold in a long semiconservative genome, while cancer cells display error rates over three orders of magnitude greater). Furthermore, the
rapid replication rates that allow such high mutation rates in the conservative
case provide no help, as the maximum allowed mutation rate cannot exceed 
a rather low threshold value, no matter how fast the cancer cells replicate. 
Hence, these recent results appear to present a paradox: rapidly mutating
genomes are prevalent in cancerous cells, but such high mutation rates 
should exceed the error threshold and hence yield inviable genomes.

In this paper, we address this paradox and demonstrate that a semiconservative
system can mimic a conservative population through the degradation of 
post-methylation lesion repair. In section II we discuss the isomorphism between
conservative replication and semiconservative replication without lesion
repair. In section III we look at the implications of this result and in section
IV we present our conclusions.

\section{DNA Repair and Semiconservative Replication}

As discussed above, DNA replication can be considered a three part process;
unzipping, complementary strand creation and mismatch repair. Afterwards,
any remaining mismatches, as well as damage caused by environmental conditions,
are repaired by a set of repair enzymes. Global genomic repair (GGR) fixes 
lesions, errors and mismatches along the entire genome, while transcription
coupled repair (TCR) subjects the expressed portion of the genome to more
careful scrutiny and repair.

In appendix A, we use the quasispecies equations to demonstrate a mathematical
isomorphism between a population of conservatively replicating 
genomes and semiconservatively replicating genomes without any lesion repair.
In this case, the semiconservative system behaves, in essence, like a
conservatively replicating system on a transformed landscape. Each single 
stranded genome produces one, possibly error prone, complementary copy. 
Although mismatch repair may keep the effective error rate low, the lack
of lesion repair ensures that the original strand is unaffected by these
errors. Hence, each genome replicates, in essence, conservatively, but with the
added wrinkle that each single stranded genome remains attached to the strand
that either created it or that it most recently created, yielding a system 
that replicates conservatively on a transformed fitness landscape.

To make this more rigorous, appendix B presents the full solution to the 
semiconservative quasispecies evolving on a specific single fitness peak landscape.
This is plotted in Fig. 3, along with the conservative and semiconservative
solutions to the same problem. 
While the semiconservative error threshold clearly plateaus at high $\sigma$, 
the repair-free semiconservative case mimics a conservative system, as the
error threshold increases indefinitely with increasing $\sigma$.

\begin{figure}[ht]
\begin{center}
\includegraphics[width=3.3in]{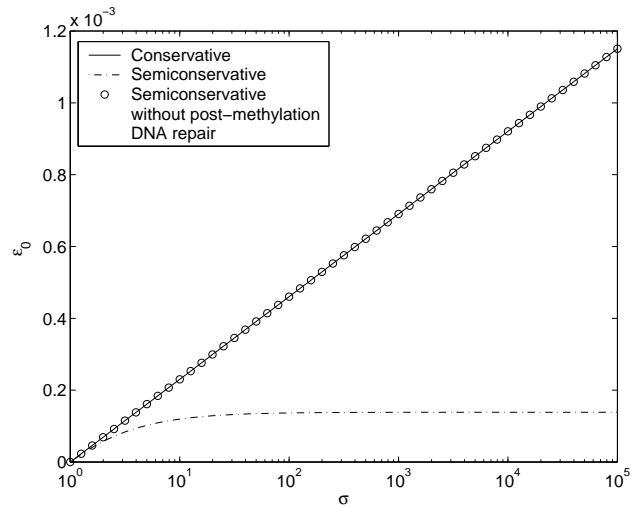}
\end{center}
\caption{The value of the error threshold ($\epsilon_0$) vs. the fitness 
of the master sequence relative to the rest of the population ($\sigma$) on 
a single fitness peak landscape. 
The genome length is set to $N = 1 \times 10^4 $.
Conservative, semiconservative and semiconservative systems without 
post-methylation lesion repair are all shown.}
\end{figure}
 Lastly, it is important to consider the case where lesion repair is
partially active, as complete degradation of lesion repair is not likely
to occur in nature. In appendix C, the single fitness peak quasispecies is
reconsidered, this time with partially active lesion repair.  
The error threshold
is shown to increase smoothly from the semiconservative to the conservative
threshold as shown in Fig. 4. This turns out to be important, and will be
discussed in
the next section.
\begin{figure}[htb]
\begin{center}
\includegraphics[width=3.3in]{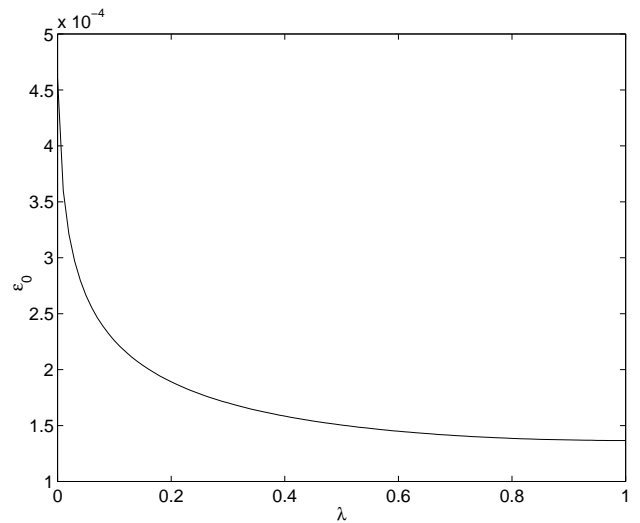}
\end{center}
\caption{The error threshold ($\epsilon_0$) for a semiconservatively replicating population
on a single fitness peak landscape vs. the probability of post-methylation
DNA repair ($\lambda$). Here, $N = 1 \times 10^4$ and $ \sigma = 1 \times 10^2 $.}
\end{figure}

\section{Discussion}

As discussed above, recent results on the semiconservative quasispecies has
presented a paradox in tumor progression. In the last section, we 
presented a possible resolution. Although extremely
high mutation rates and genetic instability are found in all cancer cells,
these mutation rates cannot be handled by a semiconservative genome. However,
as lesion repair begins to fail, the error threshold increases. 
These increased mutation rates may in turn lead to further failure of
the lesion repair system and a higher threshold, creating a positive feedback
cycle. 

The concept of a mutator phenotype in cancer has a long and established history
\cite{Loeb1}. Failure to prevent and repair mutations is well documented in cancer
cells \cite{XP3}, yet it remains unclear to what extent each repair enzyme
is or is not active in any given cell. Thus, it is difficult to say with
conviction that lesion repair failure is indeed a prerequisite for 
the sustenance of genetic instability in MIN genomes in
nature. However, there are a number of encouraging signs that this is, in fact,
the case. Many human tumors have been found to be deficient in checkpoint 
pathways, including those that involve p53, p16 and p19ARF \cite{Loeb1, Sherr}.
These checkpoints are, among other things, designed to increase the efficiency of DNA 
post-methylation damage repair, and mice that lack these checkpoint genes display
abnormally high levels of spontaneous tumor incidence \cite{Loeb1}. Loss of the
p53 tumor suppression gene has been shown to lead to less efficient GGR 
\cite{Ford1, Ford2, Ford3} and mutations of the BRCA1 gene, which enhances
the GGR process, and greatly increases the risk of breast cancer in women 
\cite{Ford4}. Well documented diseases, such as Xeroderma Pigmentosum 
\cite{XP1, XP2, XP3} are caused by defects in GGR and manifest themselves as an overwhelmingly high probability of tumor development. Expression profiles
of pancreatic cancer cells have demonstrated down-regulation of DNA 
repair genes \cite{Tatjana}. All-{\it trans}-retinoic acid has been shown to 
prevent certain carcinogenic transformations by enhancing DNA
repair through checkpoint effects \cite{Dragnev}. Lastly, numerous human studies
have shown positive correlations between individual fluctuations
in DNA repair capability and cancer risk \cite{Berwick}.

Although it is clear that DNA repair is linked to tumor suppression, we
are suggesting a fundamentally different outlook on the problem. Rather than
simply protecting the genome from mutations, DNA repair also prevents the 
proliferation of genomes with high mutation rates. As methyl-directed mismatch repair begins
to fail, the error threshold will soon be crossed, unless DNA repair begins
to fail as well. The available experimental evidence shows a definite 
correlation between repair failure and cancer risk, but causation is
not evident. As well, alternative hypotheses can explain this correlation (as
repair failure makes genetic instability more likely, and genetic instability
makes repair failure more probable), so more evidence is required.
One fundamental and novel aspect of our hypothesis, although, of course 
enormously difficult to practically implement,
is that reinstating lesion repair in a full-blown cancer cell 
should lower the error 
threshold and provide the same effect as pushing the 
cells past the error threshold, without the added side effects associated with
introducing mutagens into the body.

One must take great care in making grand statements regarding complex biological
systems from simplified physics model. Complex processes involving numerous
enzymes and prefactors are incorporated into first-order rate constants, 
and the various types of DNA damage are ignored in favor of simple point 
mutations. Further, DNA repair covers a complex set of phenomena, rather
than the simple post-methylation mismatch repair treated in the model. 
However, the quasispecies model has been impressively successful
in dealing with a wide variety of complex systems, including finite
populations \cite{Zhang, Alves, Campos}, time dependent landscapes 
\cite{timedep, timedep2, timedep3}, punctuated equilibrium \cite{Zhang, Krug}
and even the accurate prediction of human B-cell mutation rates \cite{Kamp}
and viral properties \cite{Kamp2, Brumer1}. Despite its simplicity, the model
seems to capture the robust properties of genomic evolution. Furthermore,
it is successful at all mutation rates, whereas many theories of population
genetics only work at low mutation rates, which obviously does not apply to
genetically unstable tumor progression. Regarding the fitness landscape, although cancerous 
genomes can be highly heterogeneous, the single fitness peak landscape 
likely captures the 
general features of local behavior even on more complex landscapes, and 
can be shown to yield the same behavior as more delocalized landscapes.
As well, the mathematical isomorphisms shown in the appendices hold for all
landscapes and W matrices.

Lastly, although the model is restricted to errors in the form of point mutations, 
these are the major source of genetic instability in the MIN (microsatellite
instability) tumors, which can be found in 13\% of sporadic colon cancers 
\cite{Kinzler} and
all hereditary nonpolyposis colorectal cancer. MIN tumors can display a 
point mutation rate one thousand times greater than that of a normal cell
\cite{Nowak, Cahill}.
Other tumors display
genetic instability in the form of CIN (chromosomal instability) with a wide 
variation in chromosome number and other chromosomal instability 
\cite{Cahill, XP2}. The possibility that such instability can be treated,
to a first approximation, 
through the inclusion of recombination \cite{recomb, Barnett} and simulation
techniques is the subject
of future research \cite{BrumerMichor}. Previous results in these areas provide reason to 
believe that the underlying dynamics for models of CIN tumors should provide
similar results to those obtained here. 

\section{Conclusions}

In this paper, we presented a paradox arising from recent results regarding
the quasispecies model of semiconservative replication. The relatively low
values for the error threshold, together with the fact that this threshold does
not increase with increasing master sequence fitness, suggests that a 
true semiconservative system should not be capable of handling the exceptionally
high  mutation rates associated with cancer cells. We demonstrated that,
through the degradation of lesion repair, the semiconservative system 
begins to mimic its conservative counterpart, with an increasing
error threshold whose upper bound becomes increasingly dependent on the replication rate. Thus, we postulate that the failure of mismatch repair systems
and the corresponding increase in mutation rates that are found in MIN tumors
must be accompanied by failure of post-methylation lesion repair. Although we
present some experimental evidence to support this, the simplicity of the
model together with the complexity of the problem require further experimental evidence to fully justify our claim. Thus, we have suggested a new outlook to
guide further experimentation and more complex model calculations.
 
\begin{acknowledgments}

The authors are indebted to Franziska
Michor and Emmanuel Tannenbaum for invaluable discussions and a careful reading of the manuscript.
This research was supported by an NIH postdoctoral fellowship.
\end{acknowledgments}

\appendix
\section{Semiconservative Quasispecies Model Without Post-Methylation DNA Repair}

In this appendix we examine a semiconservatively replicating quasispecies 
model in the absence of any lesion repair. The standard conservative model
describes the evolution of a set of organisms, each with a genome 
$ \phi = s_1s_2\cdots s_N $,
where each $ s_i $ represents a ``letter'' chosen from an alphabet of size 
$ S $.
The population fractions obey the set of differential equations \cite{Eigen}
\begin{equation}
\frac{dx_{\phi}}{dt} = \sum_{\phi'}A(\phi')W(\phi, \phi')x_{\phi'} - f(t)x_{\phi},
\end{equation}
where $ x_{\phi} $ denotes the fraction of the population with genome $\phi$,
$ A(\phi) $ represents the fitness, or growth rate, of sequence $\phi$,
and $W(\phi, \phi')$ is the likelihood of creating sequence $\phi$ from
$\phi'$ by mutations. $ f(t) = \sum_{\phi}A(\phi)x_{\phi} $ is the average fitness of the
population, which holds the population size constant and introduces competition.
If only point mutations are allowed and a genome-independent
mutation probability $\epsilon$ is assumed, then $ W(\phi, \phi') $
can be written in terms of the number of bases at which $\phi$ and $\phi'$
differ, the Hamming distance $ HD(\phi, \phi')$, as
\begin{equation}
W(\phi, \phi') = (\frac{\epsilon}{S - 1})^{HD(\phi, \phi')}(1 - \epsilon)^{N - HD(\phi, \phi')},
\end{equation}
where $N$ represents the length of the genome.
The isomorphism we are about to describe holds for all $ W $, but we shall limit
ourselves to this manifestation in appendix B and C.

For a semiconservative system, organisms are described by a double stranded
genome $\{\phi, \phi'\}$, the population fractions 
as $x_{\{\phi, \phi'\}}$, and the growth rates as $ A({\phi, \phi'}) $. It is 
important to note that, in the absence of lesion repair, $\phi'$ is {\em not}
defined by complementary base pairing to $\phi$, since there is no requirement
that base pair mismatches be altered. We use $W(\phi, \phi')$ as before to describe
the probability that replication of the unzipped single stranded genome 
$\phi'$ will produce new strand $\phi$.

Hence, the quasispecies equations for a semiconservative genome without 
post-methylation DNA repair can be written as
\begin{eqnarray} \nonumber
&&\frac{dx_{\phi}}{dt} = \sum_{\phi_a, \phi_b} (W(\phi, \phi_a) + W(\phi, \phi_b))A({\phi_a, \phi_b})x_{\{\phi_a, \phi_b\}} - \\ [0.2in]
&&\sum_{\phi_a} f(t) (x_{\{\phi, \phi_a\}} + x_{\{\phi_a, \phi\}}),
\end{eqnarray}
where we define the population fraction $x_{\phi} \equiv \sum_{\phi_a} 
x_{\{\phi, \phi_a\}} + x_{\{\phi_a, \phi\}} $.
Note that we count the $ 5' \rightarrow 3' $ and $3' \rightarrow 5' $ strand
separately to avoid double counting. Rearranging,
\begin{eqnarray} \nonumber
&&\frac{dx_{\phi}}{dt} = \sum_{\phi_a, \phi_b} W(\phi, \phi_a)A({\phi_a, \phi_b})x_{\{\phi_a, \phi_b\}} \\ [0.2in] 
&&+ \sum_{\phi_a, \phi_b} W(\phi, \phi_b)A({\phi_a, \phi_b})x_{\{\phi_a, \phi_b\}} - f(t)x_{\phi} \\ [0.2in] \nonumber
&& =  \sum_{\phi_a, \phi_b} W(\phi, \phi_a) (A({\phi_a, \phi_b})x_{\{\phi_a, \phi_b\}} + A({\phi_b, \phi_a})x_{\{\phi_b, \phi_a\}}) \\ [0.2in] 
&&- f(t)x_{\phi},
\end{eqnarray}
where the last expression is obtained by switching the dummy variables $\phi_a$ and
$\phi_b$ in the second summation of Eqn. (A5). We can define an average replication rate
for the single strand $\phi$ as 
\begin{equation}
\overline{A(\phi)} = \frac{\sum_{\phi_a} (A({\phi_a, \phi})x_{\{\phi_a, \phi\}} + A({\phi, \phi_a})x_{\{\phi, \phi_a\}})}{x_{\phi}}
\end{equation}
yielding the main result of this appendix,
\begin{equation}
\frac{dx_{\phi}}{dt} = \sum_{\phi'} \overline{A(\phi')}  W(\phi, \phi')x_{\phi'} - f(t) x_{\phi}.
\end{equation}
which looks remarkably similar to Eqn. (A1).
For large populations, 
$\overline{A(\phi')}$ must rapidly equilibrate and remain steady, yielding
a system of equations that are identical to that of a conservative quasispecies
with a transformed set of rate constants. This is studied in more detail in 
the particular example of appendix B.

\section{Single Fitness Peak Landscape}

In this appendix, we study the system of appendix A evolving on the commonly
used single 
fitness peak landscape described below. To solve this problem, we shall
explicitly make a number of approximations that have been well studied and
found to accurately describe the true dynamics of the system for reasonable
genome lengths and population sizes.

The single fitness peak landscape describes the situation where a specific
genome perfectly ``fits'' the environment and hence replicates rapidly, while 
all other genomes are equally poor replicators. Here, we investigate the
case where at least one strand needs to be perfect in order to be viable, 
a reasonable model for the housekeeping genes responsible for cell survival
(other landscapes will be studied in a future work \cite{BrumerManny}). 
Thus, $A(\phi_a, \phi_b) = 
\sigma \gg 1 $ if either $\phi_a$ or $\phi_b$ are in the set $\{\phi_0, \phi_0'\}$ which represent the master sequence and its perfect complement, and $A(\phi_a, \phi_b) = 1$ otherwise. 

For large populations and genome lengths, we can ignore mutations from unfit 
sequences to the master sequence (an approximation that becomes exact as the
genome length increases to infinity, but is accurate at realistic finite 
genome lengths) and assume that, at equilibrium, master genomes are paired with statistically 
distributed complements
(which is exact for this landscape in the large population limit, and
rapidly converges for finite populations).
Thus, using Eqns. (A6) and (A7) and the fact that the symmetric
equations conserve the equality of concentrations of complementary sequences, 
we can write a differential equation for the sum of the population 
fractions of the
single stranded master genome and its complement, $x_0$, and the remaining population, 
$x_1 = 1 - x_0$,
\begin{eqnarray}
&&\frac{dx_0}{dt} = q^N \sigma x_0 - f(t)x_0 \\ [0.2in] \nonumber
&& \frac{dx_1}{dt} = (1 - q^N)\sigma x_0 + \sigma x_{0,1} + x_1 - x_{0,1}\\ [0.2in]
&&- f(t)x_1
\end{eqnarray}
where $N$ is the length of the genome, $q$ represents the replicative fidelity, or $ 1 - \epsilon $, where $\epsilon$ is the per base point mutation probability, which is assumed to be sequence
independent, $ x_{0,1}$ represents the fraction of the population that are
imperfect sequences bonded to a perfect sequence (that is, the strands
that are not members of $ x_0 $, but are bonded to a member of $ x_0 $), and $ f(t) = \sigma x_0 + \sigma x_{0,1} + 1 - x_0 - x_{0,1}$. As complements
are statistically distributed, we can define $ x_{0,1} = F_{0,1}x_0 $, where
$ F_{0,1} $ represents the fraction of perfect sequences bonded to
imperfect sequences, and is independent of $ x_0 $.
We can solve these equations by searching for equilibrium solutions, 
$\dot{x}_0 = \dot{x}_1 = 0$, yielding two solutions, the quasispecies solution,
\begin{eqnarray}
&&x_0 = \frac{1 - q^N\sigma}{(1 + F_{0,1})(\sigma - 1)} 
\end{eqnarray}
and the quasispecies-free solution, $x_0 = 0$. The error
catastrophe occurs when the two solutions meet, i.e., when $ x_0 = 0 $ in the 
quasispecies solution. This gives 
\begin{equation}
\sigma = 1/q^{N},
\end{equation}
which is identical to the conservative solution and is plotted and 
discussed in section II.

\section{Degradation of Post-Methylation DNA Repair}

In this appendix, we evaluate the effect of the progressive failure of 
post-methylation DNA repair on the single fitness peak landscape discussed
in the previous appendix. Setting the
probability that an error will be repaired to be $\lambda$, we get
\begin{eqnarray} \nonumber
&&\frac{dx_{\phi}}{dt} = \sum_{\phi_a, \phi_b} (W(\phi, \phi_a, \lambda) + W(\phi, \phi_b, \lambda) + \\ [0.2in] \nonumber
&& W_{2}(\phi, \phi_a, \lambda) + W_{2}(\phi, \phi_b, \lambda) )A({\phi_a, \phi_b})x_{\{\phi_a, \phi_b\}} - \\ [0.2in] \nonumber
&&\sum_{\phi_a} f(t) (x_{\{\phi, \phi_a\}} + x_{\{\phi_a, \phi\}}) - \\ [0.2in] 
&&\sum_{\phi_a} (A({\phi, \phi_a})x_{\{\phi, \phi_a\}}  + A({\phi_a, \phi})x_{\{\phi_a, \phi\}}),
\end{eqnarray}
where $W(\phi, \phi_a, \lambda)$ represents the $\lambda$-dependent probability
that unzipped strand $\phi_a$ will produce new strand $\phi$, while 
the new quantity $W_2(\phi, \phi_a, \lambda)$ represents the probability that, after
replication and post-methylation lesion repair of unzipped strand $\phi_a$,
the erroneous repair of errors will change $\phi_a$ to strand $\phi$. 
A set of manipulations similar to those in appendix A and the definition in
Eqn. (A6) can be used to 
yield the equations
\begin{eqnarray} \nonumber
&&\frac{dx_{\phi}}{dt} = \sum_{\phi'} \overline{A(\phi')}  (W(\phi, \phi', \lambda) + W_2(\phi, \phi', \lambda))x_{\phi'} - \\ [0.2in] 
&&(f(t) + \overline{A(\phi)})x_{\phi}
\end{eqnarray}

When applied to a single fitness peak landscape, these equations can be written as 
\begin{eqnarray}\nonumber
&&\frac{dx_0}{dt} = \{(1-\frac{\lambda \epsilon}{2})^N + (1 - (1 - \frac{\lambda}{2}) \epsilon)^N\} {\sigma} x_0 \\[0.2in]
&& - (f(t) + \sigma)x_0 \\[0.2in]
&& x_1 = 1 - x_0 
\end{eqnarray}
where $x_0$ and $x_1$ and $ f(t) $ are as defined in appendix B.
This can be solved for the error threshold, which occurs when
\begin{equation}
\sigma = \frac{1}{(1 - \lambda \epsilon/2)^N + (1 - (1 - \lambda/2)\epsilon)^N - 1}.
\end{equation}
This expression is plotted and discussed in section II, and approaches the full
semiconservative treatment and the solution of appendix B in the limits 
$\lambda \rightarrow 1 $ and $\lambda \rightarrow 0$, respectively. 

\bibliography{Lesion.bib}

\end{document}